\def\arc{$\alpha$-RuCl$_3$}
\documentclass[aps,prl,twocolumn,superscriptaddress,amsmath,amssymb]{revtex4-1}
\usepackage{graphicx}%
\usepackage{textcomp}
\usepackage{color}%

\usepackage{pdfpages}
\makeatletter
\AtBeginDocument{\let\LS@rot\@undefined}
\makeatother

\begin{document}
\title{Electronic properties of \arc\ in proximity to graphene}
    
    \author{Sananda Biswas}
    \affiliation{Institut f\"{u}r Theoretische Physik, Goethe-Universit\"{a}t Frankfurt, 60438 Frankfurt am Main, Germany}
    \author{Ying Li}
    \affiliation{Institut f\"{u}r Theoretische Physik, Goethe-Universit\"{a}t Frankfurt, 60438 Frankfurt am Main, Germany}
    \affiliation{Department of Applied Physics and MOE Key Laboratory for Nonequilibrium Synthesis and Modulation of Condensed Matter, School of Science, Xi\textquotesingle an  Jiaotong University, Xi\textquotesingle an 710049, China}
    \author{Stephen M. Winter}
    \affiliation{Institut f\"{u}r Theoretische Physik, Goethe-Universit\"{a}t Frankfurt, 60438 Frankfurt am Main, Germany}
    \author{Johannes Knolle}
    \affiliation{Blackett Laboratory, Imperial College London, London SW7 2AZ, United Kingdom}
    \affiliation{Department of Physics, T33, Technische Universit\"{a}t M\"{u}nchen, 85748 Garching, Germany}
    \affiliation{Munich Center for Quantum Science and Technology (MCQST), 80799 Munich, Germany}
    \author{Roser Valent\'{\i}}
    \affiliation{Institut f\"{u}r Theoretische Physik, Goethe-Universit\"{a}t Frankfurt, 60438 Frankfurt am Main, Germany}
    
    \date{\today}

\begin{abstract} 
In the pursuit of developing routes to enhance magnetic Kitaev interactions in \arc{}, as well as probing doping effects, we investigate the electronic properties of \arc{} in proximity to graphene. We study \arc/graphene heterostructures via {\it ab initio} density functional theory calculations, Wannier projection and non-perturbative exact diagonalization methods. We show that \arc\ becomes strained when placed on graphene and charge transfer occurs between the two layers, making \arc\ (graphene) lightly electron-doped (hole-doped). This gives rise to an insulator to metal transition in \arc\ with the Fermi energy located close to the bottom of the upper Hubbard band of the $t_{2g}$ manifold. These results suggest the possibility of realizing metallic and even exotic superconducting states. Moreover, we show that in the strained  \arc\ monolayer the Kitaev interactions are enhanced by more than 50\% compared to the unstrained bulk structure. Finally, we discuss scenarios related to transport experiments in \arc/graphene heterostructures.	
\end{abstract}

\maketitle

\section{Introduction}

{\it Introduction.-} A major step towards a realization of a fault-tolerant quantum computer would be, for instance, to find materials that support bond-dependent Kitaev interactions~\cite{Kitaev1, Kitaev2} leading to a quantum spin liquid (QSL) \cite{qslBalents2017, RevModPhys.88.041002,ZhouRMP2017, qslKnolle2019} ground state of itinerant Majorana fermions that couple to static Z$_2$ gauge fields. \arc\  has been intensively discussed as a possible candidate for Kitaev physics~\cite{Sanilands15,Kim15,Banerjee16,PhysRevB.93.214431,Kim16b,rau2016spin,winter2017models,hermanns2017physics,BanerjeeQM2018, naglerNatRevPhys2019}, however it orders antiferromagnetically at low temperatures~\cite{PhysRevB.92.235119,Banerjee16} due to the presence of additional magnetic couplings~\cite{Sandilands16a,PhysRevB.93.214431, Yadav16, winter2017models} extending beyond the pure Kitaev interaction. In order to suppress magnetism and/or enhance pure Kitaev interactions in \arc, various routes are currently being pursued including the application of magnetic fields, external pressure and chemical doping~\cite{PhysRevB.92.235119, Yadav16, Baek17, Banerjee17, HentrichPRL2018, KoitzschPRM2017,WinterPRL2018, KiraPRL2019, HentrichPRL2018,Wang17b,Cui17}. Here we discuss yet another engineering route: electronic modification of \arc\ due to the proximity to graphene. We show that the formerly  Mott insulating \arc\ becomes electron-doped via a charge transfer from graphene, and that the substrate-induced strain leads to a significant enhancement of the Kitaev exchange.

In bulk $\alpha$-RuCl$_3$, the Ru atoms have a $d^5$ electronic configuration with one hole per site occupying a spin-orbital coupled $j_{\rm eff} = 1/2$ state. These holes are localized due to strong Coulomb interactions favoring a spin-orbit assisted Mott insulating phase. Due to the specific spin-orbital composition of the $j_{\rm eff}$ states, the effective magnetic couplings between the localized holes are strongly anisotropic, and can be described by extended Kitaev models~\cite{Jackeli2009, Rau14,kimPRB2014, kimPRB2012,PhysRevB.93.214431,winter2017models}. The underlying magnetic couplings are however somewhat far away from the ideal Kitaev point with its desired QSL phase, requiring some structural manipulation to yield more ideal couplings. In the latter case, physical pressure has proved to be inappropriate, as it leads to structural distortions (dimerization) that quench completely the desired properties~\cite{biesner18,HentrichPRL2018,Wang17b,Cui17}.

\begin{figure}[!b]
\centering
\includegraphics[width=0.85\columnwidth]{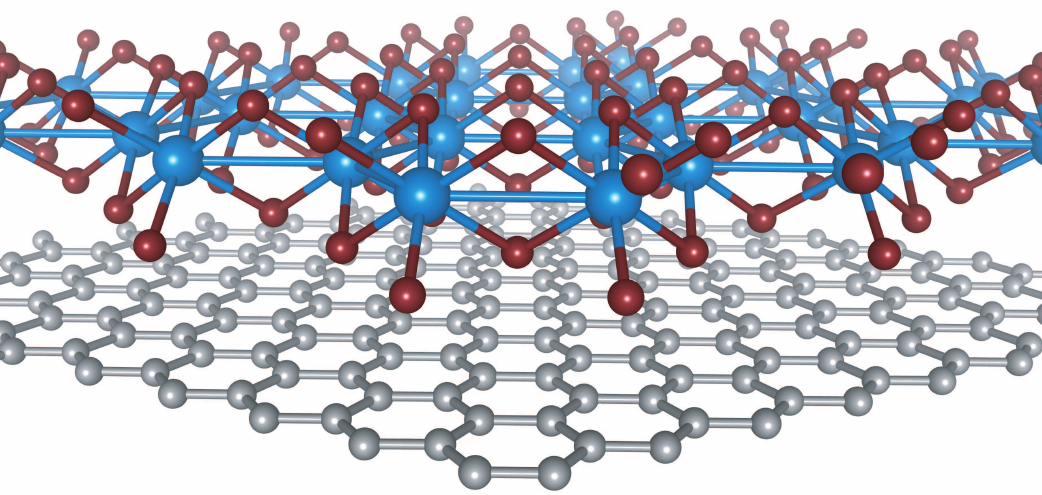}
\caption{Van der Waals bilayer \arc/graphene in the hexagonal supercell (see text for description). The blue, maroon and grey spheres represent Ru, Cl and C atoms, respectively.}
\label{fig1}
\end{figure}

On the other hand, the possibility of lightly charge doping  $\alpha$-RuCl$_3$ appears attractive, as doped Heisenberg-Kitaev models are thought to host a variety of exotic superconducting states~\cite{youPRB2012,HyartPRB2012,OkamotoPRB2013,ChoiPhysRevB2018}. At low energies, such doped materials, having an average of $(1-\delta)$ holes per Ru atom, can be described by  $tJ$-like models, including both the hopping of the excess charges and magnetic couplings between singly occupied sites. Doped materials with magnetic couplings close to the pure ferromagnetic Kitaev model are particularly attractive as  potential hosts for topological superconductivity~\cite{SchmidtPRB2018}. However, thus far, controlled doping has not yet been achieved. The authors of Ref.~\cite{KoitzschPRM2017} found that potassium doped K$_{0.5}$RuCl$_3$ (with a relatively large $\delta = 0.5$) remains insulating and appears to have a charge-ordered ground state. In the present work we argue that substrate-RuCl$_3$ heterostructures might be able to accomplish all desired features, in principle: small, controlled doping, and stable structural tuning through interface strain.

 \begin{figure}[!b]
\centering
\includegraphics[width=0.95\columnwidth]{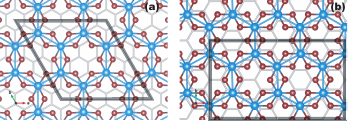}\hspace{3mm}
\caption{(a)~Top views of two supercells of \arc/gr considered in this study (black lines): (a) hexagonal and (b) rectangular. The blue and maroon spheres represent Ru and Cl atoms, respectively. The grey hexagon indicates the graphene monolayer.}
\label{fig:2}
\end{figure}

Monolayers of \arc\ have been previously fabricated through the exfoliation technique either in restacked geometry or on Si/SiO$_2$ substrates~\cite{WeberNL2016, ZhouJPCS2019}. In this paper, we have considered graphene (gr) as substrate, as it has been employed in several recent studies~\cite{zhouArXiv18, mashhadiArXiv19}. We perform {\it ab-initio} density functional theory (DFT) calculations and investigate the structural, electronic and magnetic properties of \arc\ on graphene (see Fig.~\ref{fig1}). We find that (i) \arc\ becomes strained when placed on graphene and there is a charge transfer from graphene to \arc\ making \arc\ (graphene) lightly electron-doped (hole-doped), (ii) our magnetic calculation of the strained monolayer \arc\ shows enhancement of the Kitaev interaction by more than 50$\%$  compared to the bulk \arc{} and, (iii) our electronic structure calculations suggest two alternative scenarios to interpret recent transport measurements of \arc/gr heterostructures~\cite{zhouArXiv18, mashhadiArXiv19} which could be distinguished via application of an in-plane magnetic field.

{\it Methods.-} We have performed  DFT structural relaxations of  \arc/gr heterostructures with the projector augmented wave method~\cite{Bloechl1994} using the Vienna $Ab-initio$ Simulation Package (VASP)~\cite{Kresse}. For the relaxations we considered as exchange-correlation functional the generalized gradient approximation (GGA)  including a $U$ correction for Ru $d$ orbitals, as implemented in GGA+$U$ (with $U$=1.5 eV)~\cite{Dudarev}. Electronic structure calculations were performed with various functionals; GGA, GGA+SOC (spin-orbit) and GGA+SOC+$U$ with and without inclusion of magnetism. We double-checked our calculations with the full-potential-linear-augmented-wave basis (LAPW) as implemented in the WIEN2k code~\cite{Blaha2001}.  Hopping integrals were obtained by the Wannier projector method~\cite{Foyevtsova2013, Aichhorn2009, Ferber2014} on the FPLAW results~\cite{Supplement} and the exchange parameters were estimated using the projection and exact diagonalization method of Refs.~\cite{PhysRevB.93.214431, RiedlPssb}. Charge transfer values were obtained by Bader analysis on the VASP results~\cite{Bader}.

{\it Results.-} At ambient pressure, bulk \arc\ has been reported to have either $C2/m$ or $R\bar{3}$ symmetry ~\cite{PhysRevB.92.235119,reschke2018sub,Mai2019}. While the latter case has perfect Ru hexagons, in the former case Ru hexagons exhibit a small bond-disproportionation ($l_l/l_s=$ 1.05; where $l_l$ and $l_s$ are the long and short Ru-Ru bonds, respectively). Owing to the lattice mismatch (15-17\%) between \arc\ (5.80 \AA; considering in-plane C$_3$ symmetry) and graphene (2.46 \AA), we considered  two different heterostructure supercells for \arc/gr (see Fig.~\ref{fig:2}): (a) an hexagonal supercell  containing 82 atoms (composed of a 5$\times$5 graphene supercell and a $\sqrt{3}\times\sqrt{3}$ \arc\ monolayer) and (b) a rectangular supercell containing 112 atoms. In both these supercells,  due to the strong carbon $sp^2$ bonding, the graphene layer is kept unstrained i.e., all C-C bond-lengths are 1.42 \AA. By keeping the lattice parameters fixed, we performed ionic relaxations of the RuCl$_3$ layer within spin-polarized DFT in the GGA approximation. We included van der Waals corrections. The resulting relaxed structures are dependent on the relative stacking of the layers. In the hexagonal supercell (Fig.~\ref{fig:2}(a)), each Ru hexagon is undistorted, i.e., no bond-disproportionation is observed. The \arc\ layer develops a positive (expansive) strain (2.5\% tensile)~\cite{caldetails} compared to the corresponding bulk structure. On the other hand, in the rectangular supercell Fig.~\ref{fig:2}(b), the Ru-Ru bonds are slightly anisotropic (bond disproportionation $l_l/l_s=$ 1.05) and the \arc\ layer mimics scenarios corresponding to negative (compressive) strain (-5\% tensile). Note that this bond-disproportionation is comparable to the bulk ambient pressure $C2/m$ structure, and much smaller than the dimerized high-pressure structures~\cite{biesner18,HentrichPRL2018}. The van der Waals distance between graphene and \arc\ is 3.37 \AA\ in both supercells.

 \begin{figure}[t!]
\centering
\includegraphics[width=0.85\columnwidth]{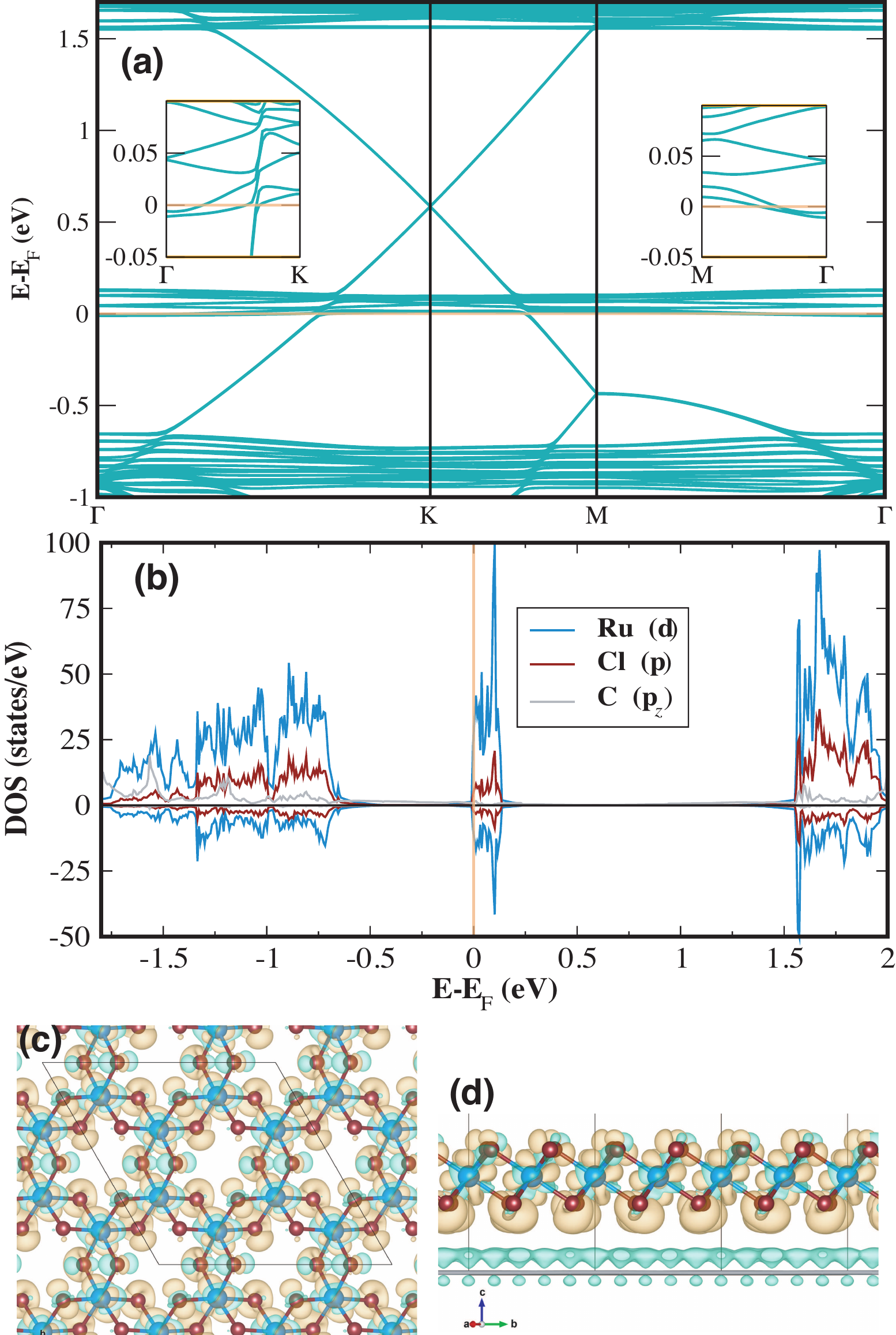}\hspace{3mm}
\caption{(a) Calculated band structure along the high-symmetry points ($\Gamma$, K, M) of the hexagonal Brillouin zone for zigzag antiferromagnetic \arc/gr in the GGA+SOC+$U$ scheme with $U=$1.5 eV. The insets show zoomed in regions near the Fermi level, $E_F$. (b) Atom projected spin-polarized density of states. The $E_F$ lies slightly above of the upper Hubbard band in the $t_{2g}$ manifold. (c)-(d) Top of side view of charge difference, $\delta \rho $ plot with charge isosurface $5 \times 10^{-4}$ $e$/\AA$^{3}$. The peach and cyan colors represent regions with charge accumulation and depletion (deficiency), respectively. See text for the definition of $\Delta \rho$. }
\label{fig:3}
\end{figure}
 
In Figs.~\ref{fig:3}(a)-(b), we show the band structure and density of states (DOS)  obtained within the GGA+SOC+$U$ approximation ($U=$ 1.5 eV) for the relaxed hexagonal supercell structure in the zigzag antiferromangetic configuration. Here, the Ru spin-orbit entangled $j_{\rm eff}=1/2$ and $j_{\rm eff}=3/2$ manifolds split into upper and lower Hubbard bands (splitting $\approx$ 0.5-0.6 eV for the $U$ value considered in the calculations), respectively. The Dirac cone of graphene is shifted up from the Fermi energy $E_F$ (compared to the bare graphene) by $\approx$ 0.7 eV,  indicating a charge transfer from graphene to \arc. As a result, $E_F$ lies at the bottom of the rather flat upper Hubbard band of \arc. In the insets of Fig.~\ref{fig:3}(a), we observe that the hybridization occurs between \arc\ and graphene only in a small region of the Brillouin zone. Except for these points, the \arc\ bands remain almost flat.  Comparison with non-magnetic calculations~\cite{Supplement} show only small modifications at the hybridization region.

In order to estimate the degree of charge transfer, we computed from DFT wavefunctions (within VASP) the Bader charges of the heterostructure. For the hexagonal supercell, the amount of charge transfer is $\delta = 0.064$ $e$ per RuCl$_3$ unit (-0.010 $e$ per carbon). Thus, graphene becomes electron deficient and \arc\ becomes electron-rich. Note that positive (negative) values of $\delta q$ refer to charge accumulation (depletion).  To visualize the charge transfer, in Fig.~\ref{fig:3}(c)-(d) we have plotted, for a particular isosurface value, the charge density difference $\Delta \rho = \rho({\rm gr/RuCl_3}) - \rho({\rm gr}) - \rho({\rm RuCl_3})$, where $\rho$ is the charge density in $e$/\AA$^3$. As expected from the charge transfer values,  the charge deficiency occurs at the graphene layer with the transferred charge accumulating mostly around the Cl atoms of \arc\ positioned close to graphene. Interestingly, the charge distribution around Ru and the Cl atoms away from the graphene layer shows regions of charge depletion and regions of charge accumulation following the Ru and Cl electronegativity differences (see the charge-difference side-view displayed in Fig.~\ref{fig:3}(d)). Calculations for the rectangular supercell show essentially the same degree of charge transfer, suggesting that the doping level is not strongly stacking-dependent~\cite{Supplement}.

In principle, the low-energy Hamiltonian for an electron-doped layer of $\alpha$-RuCl$_3$  can be expressed in terms of a $tJ$-like model with anisotropic magnetic couplings between nearest-neighbor sites $i$ and $j$:
\begin{align}\label{eq:1}
\mathcal{H} = \sum_{\langle ij \rangle} t_{ij} \mathbf{c}_{i}^\dagger \mathbf{c}_j + \mathcal{H}_{ij}^{\rm spin}
\end{align}
where $\mathbf{c}_i^\dagger = (c_{i,\uparrow}^\dagger \ c_{i,\downarrow}^\dagger)$ creates an electron in the $j_{\rm eff}$ = 1/2 state at site $i$, $t_{ij}$ is the hopping integral between sites $i$ and $j$ and $\mathcal{H}_{ij}^{\rm spin}$ describes  an extended Kitaev model of the form:
\begin{align}\label{eq:hamil}
\mathcal{H}_{ij}^{\rm spin} & =  J_{ij} \mathbf{S}_i \cdot \mathbf{S}_j + K_{ij}  S{_i^\gamma} S{_j^\gamma} + \Gamma_{ij} (S{_i^\alpha} S{_j^\beta} +S{_i^\beta} S{_j^\alpha}) \\ \nonumber
&+ \Gamma_{ij}^{\prime} (S{_i^\gamma} S{_j^\alpha} +S{_i^\gamma} S{_j^\beta} + S{_i^\alpha} S{_j^\gamma} +S{_i^\beta} S{_j^\gamma}),
\end{align}
where  $\mathbf{S}$  corresponds to the $j_{\rm eff}$ = 1/2 operator with $\alpha,\beta,\gamma = x,y,z$. In order to estimate the hopping and interaction parameters of $\mathcal{H}$, we performed additional calculations, as described above, on isolated and charge-neutral layers of $\alpha$-RuCl$_3$ employing the relaxed hexagonal geometry of the Ru layer in the $\alpha$-RuCl$_3$/gr heterostructure (Fig.~\ref{fig:2}(a)). These calculations are very useful to identify the position of  \arc/gr  in the phase diagram of the doped Kitaev-Heisenberg model away from half-filling~\cite{youPRB2012, HyartPRB2012, OkamotoPRB2013, ChoiPhysRevB2018}. The hopping between nearest neighbor sites is given by $t_{0} = \frac{1}{3}\left( 2t_{1} +t_3\right)$, in terms of the hopping integrals defined in Ref.~\cite{PhysRevB.93.214431}. Our estimates via Wannier projection of the LAPW bandstructure show that this value is very small, $t_0\sim 7$ meV, suggesting that the hopping between $j_{\rm eff} = 1/2$ states is rather suppressed. Table~\ref{table_mag} displays the values of the nearest-neighbor magnetic couplings  in the  strained \arc\ monolayer estimated from exact diagonalization of the {\it ab initio}-derived multiorbital Hubbard model on two-site clusters~\cite{PhysRevB.93.214431,Supplement}. Comparing the values with those of  bulk \arc~\cite{PhysRevB.93.214431}, we find that $|K|$  increases by more than 50\% with respect to bulk $\alpha$-RuCl$_3$, while $|J|$ and $|\Gamma|$ terms decrease and  $|\Gamma|^{\prime}$ remains almost unchanged. Note that the case studied here is different from that in Ref.~\cite{gerberArXiv19}, where no strain effects are taken into account.

\begin{table}[t!]
\caption{ \label{table_mag} Comparison of magnetic interactions in meV for strained \arc\ (see text for description) from current study and unstrained  Z-bond of bulk \arc\ in $C/2m$ structure from Ref.~\cite{PhysRevB.93.214431}. Values are obtained by exact diagonalization on two-site clusters employing $U$ = 3 eV, $J_H$ = 0.6 eV, $\lambda$ = 0.15 eV.}
\begin{ruledtabular}
\begin{tabular}{ccccccc}
&Bond & $J$ & $K$ & $\Gamma$ & $\Gamma^{\prime}$ & $|K/J|$\\
\hline  \\[-2.0ex]
&X, Y       & -0.5 & -16.8  & +1.8 & -2.7 & 33.60\\
&Z          & -0.4 & -17.2  & +1.9 & -2.4 & 43.00\\
&Z ($C2/m$) & -3.0 &  -7.3  & +8.4 & -2.0 &  2.43\\
\end{tabular}
\end{ruledtabular}
\end{table}

These changes can be correlated with the structural changes: \arc/gr exhibits a larger Ru-Cl-Ru bond angle (96.54$^\circ$) compared to the ambient condition bulk structure (94$^\circ$). This has the effect of significantly suppressing the direct Ru-Ru hopping. Instead, the dominant hopping occurs via hybridization with the Cl ligands, as considered by the original proposal of Ref.~\cite{Jackeli2009}. Since the direct Ru-Ru hopping is simultaneously the source of the non-Kitaev magnetic couplings and the $t_0$ hopping in Eq.~(\ref{eq:1}), both are suppressed in the strained RuCl$_3$ layer.

{\it Discussion.-} The phase diagram of the doped Kitaev-Heisenberg model, as suggested by Refs.~\cite{youPRB2012, HyartPRB2012,OkamotoPRB2013} reveals a spin-triplet $p$-wave superconductor at low values of $J_{ij}/t_{0}$, while at higher values the stable phases are spin-singlet $s$- and $d$-wave superconductors. A nontrivial topological $p$-wave superconducting phase is reported to exist in a region between hole doping parameters (1-$\delta$) = $0.25-0.4$, when the condition, $K=t_{0}$,  is fulfilled. \arc/gr is close to satisfying this criterion ($K \approx 17$ eV and $t_0 \approx$ 7 meV). Assuming the particle-hole symmetry remains preserved, from the doped phase diagram of Refs.~\cite{youPRB2012, HyartPRB2012,OkamotoPRB2013}, one can conclude that \arc/gr shows the possibility to exhibit a trivial $p$-wave superconducting state. We note, however, that $t_{0}$  is very small so that this region may be difficult to access due to the low mobility of the doped charges.

In this context, we also investigated another heterostructure system: \arc\ on hexagonal boron-nitride (\arc/h-BN). Due to the presence of the semiconducting substrate, the amount of charge transfer decreases compared to the \arc/gr case, the values being $\delta = 0.011$ $e$ per RuCl$_3$ unit (-0.003 $e$ per BN unit). An enhanced doping can be realized in both \arc/gr and \arc/h-BN by the application of gate-voltage that may shift these heterostructures into a non-trivial topological superconducting regime. Thus, both systems have the potential to host  trivial and non-trivial topological spin-triplet superconducting states. Furthermore, we note, by analyzing Table~\ref{table_mag}, that the strained undoped \arc\ geometry satisfies the condition of $|K/J| > 8$ which is the topologically interesting region where the Kitaev QSL phase exists. This study emphasizes the high sensitivity of the properties of \arc\ on the substrate that is being chosen.

Focussing now on the interplay of itinerancy and magnetism in \arc/gr, two very recent experiments have reported transport measurements of \arc/gr heterostructures~\cite{zhouArXiv18, mashhadiArXiv19}. While both of these show evidence of charge transfer, they differ in several respects. The authors of Ref.~\cite{zhouArXiv18} observe a clear transport anomaly around a temperature of about 20 K attributed to a magnetic transition. However, the precise role of magnetism in their set-up is unclear due to a possible inhomogeneous interface between the two layers. In contrast, Ref.~\cite{mashhadiArXiv19} uses encapsulated heterostructures and a similar anomaly in transport is absent. However, the latter authors observe clear Shubnikov de Haas oscillations providing clear evidence of charge transfer. These oscillations have an unusual non-Lifshitz Kosevich (non-LK) temperature dependence with a maximum at a temperature close to the bulk $T_N$ of \arc\  which is taken as evidence of spin fluctuation-mediated electron transport scattering arising from an underlying magnetic transition. An alternative explanation could arise from the hybridization of the doped itinerant graphene band structure and the almost flat upper Hubbard band of \arc{}, which is  similar to scenarios of anomalous quantum oscillation (QO) as discussed for SmB$_6$~\cite{KnollePRL2015, ZhangPhysRevLett2016}. There, a maximum in the amplitude occurs at a temperature scale of the hybridization giving rise to a similar non-LK dependence. It will be important to distinguish the scenario of spin scattering versus anomalous QOs  for example via the application of an in-plane field which is known to suppress  the magnetic phase in \arc\ \cite{PhysRevB.92.235119} which would change the magnetic scattering but not the hybridization.

{\it Conclusions.-} We have studied the electronic and magnetic properties of \arc/gr heterostructures by a combination of  {\it ab initio} density functional theory calculations, Wannier projection and exact diagonalization of finite clusters. Our results show that  \arc\ in the \arc/gr heterostructure gets strained due to the lattice mismatch between  \arc\ and graphene and there is a charge transfer between the two layers making \arc\ (graphene) lightly electron-doped (hole-doped). Recent experimental realizations of such heterostructures~\cite{zhouArXiv18, mashhadiArXiv19} confirmed the charge transfer character found in our calculations and we proposed  measurements under in-plane magnetic field to disentangle the interplay of itinerancy and magnetism in the heterostructures.

Calculation of the hopping and exchange interaction parameters of a putative $tJ$ model would place this system in a region of possible $p$-wave supperconductivity. Furthermore, the strained monolayer \arc\ shows an enhancement of the Kitaev interaction by a factor of two with respect to bulk \arc. This suggests that by means of making the \arc\ monolayer charge-neutral in such a heterostructure geometry, for instance by incorporating a spacing layer and/or introducing absorbates to saturate graphene  would thus potentially bring this system close to the classified Kitaev QSL phase. In conclusion, the \arc/gr system lies on the boundary of a myriad of applications with high tunability for exploring exotic phases and possible technological applications.

\begin{acknowledgements}
	We thank Erik A. Henriksen and A. W. Tsen for useful discussion.
	This project was supported by the Deutsche Forschungsgemeinschaft (DFG) through grant VA117/15-1. 
\end{acknowledgements}

\bibliography{RuCl+gr}

\begin{thebibliography}{57}
\bibitem[{cal()}]{caldetails}
\bibinfo{note}{The strains are calculated with respect to the ideal bond length
  in the hexagonal $\alpha$RuCl$_3$ in order the compare the values in the two
  supercell on the same ground}.
\end{thebibliography}
\clearpage


\section*{Supplementary material}

# Electronic properties of $\alpha$-RuCl$_3$ in proximity to graphene

Sananda Biswas,[1] Ying Li,[1,2] Stephen M. Winter,[1] Johannes Knolle,[3,4,5] and Roser Valentí[1]

[1]*Institut für Theoretische Physik, Goethe-Universität Frankfurt, 60438 Frankfurt am Main, Germany*
[2]*Department of Applied Physics and MOE Key Laboratory for Nonequilibrium Synthesis and Modulation of Condensed Matter, School of Science, Xi'an Jiaotong University, Xi'an 710049, China*
[3]*Blackett Laboratory, Imperial College London, London SW7 2AZ, United Kingdom*
[4]*Department of Physics, T33, Technische Universität München, 85748 Garching, Germany*
[5]*Munich Center for Quantum Science and Technology (MCQST), 80799 Munich, Germany*

(Dated: August 26, 2019)

# Supplemental Material

## A. Methods of calculation

The structural relaxation and the related post-processing calculations (except the calculation of hopping parameters) were performed using *ab initio* DFT as implemented in the VASP package [1]. We used the projector augmented plane-wave (PAW) basis sets and the corresponding cutoff was set as 600 eV. The generalized gradient approximation (GGA) to the exchange correlation functional was used with and without $U$ correlation correction as impletemented by Dudarev *et al* [2]. The value of $U$ (=1.5 eV) was chosen to be same as Ref. [3] Van der Waals forces are included through the DFT-D2 scheme of Grimme [4]. In order to mimic the periodic boundary conditions perpendicular to the heterostructure, a vacuum separation of 20 Å is used between the periodic images.

In order to perform structural optimization, the Ru atoms were allowed to have long-range ferromagnetic configuration within GGA. Energetics of the final geometry in ferromagnetic (FM), zigzag antiferromagnetic (zzAFM) and nonmagnetic (NM) configurations, show that the NM configuration to be higher in energy than the two magnetic configurations, whereas the FM and zzAFM configurations are almost degenerate. Note that we have restricted ourselves to collinear magnetic structures for the ease of calculation and therefore, the current study does not rule out the possibility of having a noncollinear magnetic groundstate structure at low temperatures.

It is worthwhile to mention here that the initial self-consistent calculation of the final optimized structure for the zzAFM configuration within the GGA+SOC+$U$ resulted in a final ferromagnetic moments. Therefore, the constrained magnetic calculations are performed by introducing a penalty function [5] to constrain the direction of magnetic moments parallel to lattice parameter the $a$; convergence with respect to the penalty parameter is achieved to obtain the final result.

In order to calculate the amount of charge transfer between the layers, we have used Bader analysis [6] of wavefunctions obtained from VASP calculations. Finally, the hopping parameters are obtained using the WIEN2k code as described below (see section E).

Table I: Relaxed geometries of the $\alpha$-RuCl$_3$ layer in both supercells for the $\alpha$-RuCl$_3$/gr system and in the hexagonal supercell for $\alpha$-RuCl$_3$/h-BN system. The in-plane lattice constants for the rectangular supercell are $a$=10.078 Å and $b$=9.843 Å, while the values for the hexagonal supercell are $a$=$b$=12.300 Å.

| Bond-lengths | Hexagonal supercell | Rectangular supercell |
|---|---|---|
| $\alpha$-RuCl$_3$/gr | | |
| Ru-Ru | 3.55 Å | 3.17-3.18 Å ($l_s$) |
| | | 3.33-3.35 Å ($l_l$) |
| Ru-Cl | 2.39-2.40 Å | 2.34-2.36 Å |
| ∠Ru-Cl-Ru | 96.92° | 85.02° -90.67° |
| $\alpha$-RuCl$_3$/h-BN | | |
| Ru-Ru | 3.54-3.56 Å | - |
| Ru-Cl | 2.37-2.38 Å | - |
| ∠Ru-Cl-Ru | 96.21° -97.55° | - |

## B. Geometrical optimization and choice of the commensurate supercells

For the $\alpha$-RuCl$_3$/gr heterostructure, we have chosen two commensurate supercells: (a) a hexagonal supercell with 82 atoms and a rectangular supercell with 112 atoms. The k-point meshes of size $6\times6\times1$ and $2\times6\times1$ (with smearing width of 0.01) were used for the hexagonal and rectangular supercells, respectively. All the atoms in the supercell were relaxed until forces on each atoms were less than 0.002 eV/Å. The lattice parameters for the two supercells were chosen in such a way that the graphene layer is not strained (i.e., C-C bond-length remains 1.42 Å after relaxation).

While no buckling of the graphene is observed, buck-

erostructure system and $\rho$(RuCl3) [$\rho$(gr)] is the charge density of the $\alpha$-RuCl$_3$(gr) in the hetetrostructure geometry with graphene ($\alpha$-RuCl$_3$) layer removed. The unit of charge density $\rho$, is $e/\text{Å}^3$.

### D. Band structure of $\alpha$-RuCl$_3$/gr: effect of magnetism, correlation and spin-orbit coupling

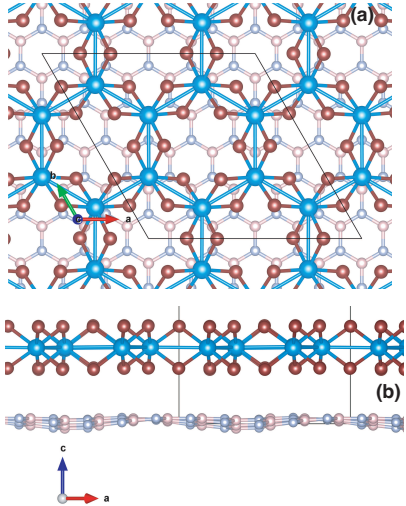

Figure 1: Top (a) and side (b) views of the $\alpha$-RuCl$_3$/h-BN system in the relaxed geometry using the hexagonal supercell. The buckling of the h-BN layer is clearly shown in the side view. Pink and violet colors represent boron and nitrogen atoms, respectively.

ling of $\approx 0.5$ Å is observed for the h-BN layer in the $\alpha$-RuCl$_3$/h-BN heterostructure (see Fig. 1) as a result of using the same lattice parameters as in the $\alpha$-RuCl$_3$/gr calculation; note that the $\alpha$-RuCl$_3$ layer remains flat after the relaxation. The structural details and different bond-lengths for both $\alpha$-RuCl$_3$/gr and $\alpha$-RuCl$_3$/h-BN are shown in Table I.

Strain values that are reported in the main text, are calculated using the formula: $\frac{d - d_\alpha}{d_\alpha}$; where $d$ is the average Ru-Ru bond-length of $\alpha$-RuCl$_3$ layer in the heterostructure and $d_\alpha$ is the same for the bulk $\alpha$-RuCl$_3$ (taken to be 3.45 Å here).

### C. Charge transfer

As mentioned above, the amount of charge transfer was obtained by Bader analysis [6]. Here, we show that the value of the charge transfer remains almost similar for different magnetic configurations. We note here that using the rectangular supercell, we obtain only a slightly lower value of $\delta$ (0.050 per RuCl$_3$) compared to the hexagonal supercell; this can be attributed to the larger strain value in the former case and slightly inhomogeneous Ru-Ru bond-lengths. Table II shows that the values of the charge transfer are almost unchanged in the rectangular supercell for different magnetic configurations within GGA, GGA+$U$ and GGA+SOC+$U$.

As mentioned above, the amount of charge transfer was obtained by Bader analysis [6]. The charge difference of a heterostructure configuration is given by $\Delta \rho = \rho(\text{gr/RuCl}_3) - \rho(\text{gr}) - \rho(\text{RuCl}_3)$, where the $\rho(\text{gr/RuCl}_3)$ is the total charge density of the het-

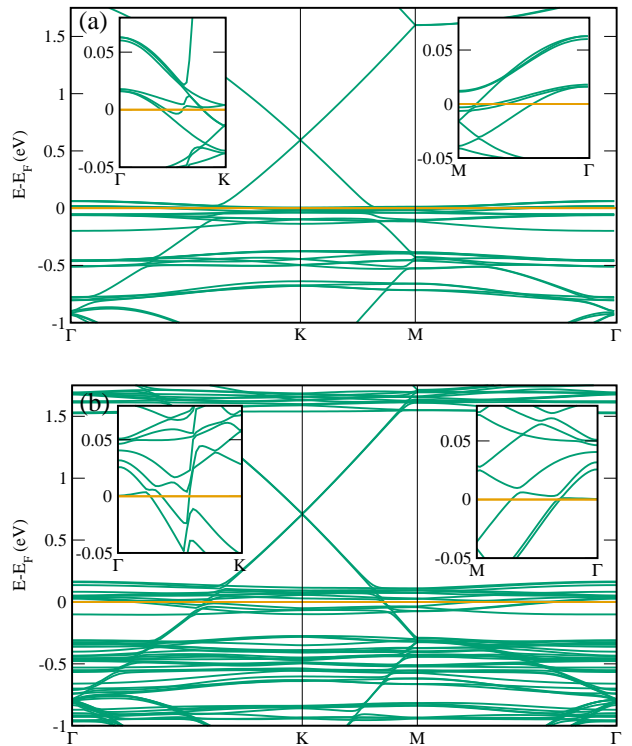

Figure 2: Non-spin-polarized band structure within (a) GGA and (b) GGA+SOC schemes for $\alpha$-RuCl$_3$/gr in hexagonal supercell. Insets show zoomed-in regions of the band structures around the Fermi level, $E_F$.

Here we compare the non-spin-polarized band structure within the GGA and GGA+SOC schemes for $\alpha$-RuCl$_3$/gr system (see Fig. 2). We find that the spin-orbit coupling splits the bands and this changes the details of the hybridization. Therefore, the Fermi energy $E_F$, shifts down in Fig. 2(b) compared to Fig. 2(a). One can compare Fig. 2(b) with Fig.3(a) of the main text, in order to observe the role played by magnetism and correlation; magnetism modifies the hydridization and correlation increases the splitting between the upper and lower Hubbard bands.

We performed additional band structure calculations in the zigzag antiferromagnetic configuration with the WIEN2k code for the $\alpha$-RuCl$_3$/gr heterostructure with two magnetization directions within the GGA+SOC+$U$

Table II: Amount of charge transfer in the rectangular supercell for different magnetic structures within different approximations.

| Calculation method | Nonmagnetic | Ferromagnetic | Zigzag antiferromagnetic |
|---|---|---|---|
| GGA | -0.0087/C <br> 0.046/$\alpha$-RuCl$_3$ | -0.0086/C <br> 0.046/$\alpha$-RuCl$_3$ | -0.0088/C <br> 0.047/$\alpha$-RuCl$_3$ |
| GGA+$U$ | -0.0110/C <br> 0.059/$\alpha$-RuCl$_3$ | -0.0076/C <br> 0.041/$\alpha$-RuCl$_3$ | -0.0094/C <br> 0.050/$\alpha$-RuCl$_3$ |
| GGA+SOC+$U$ | -0.0091/C <br> 0.049/$\alpha$-RuCl$_3$ | - | -0.0095/C <br> 0.050/$\alpha$-RuCl$_3$ |

scheme: one in-plane direction along $a$ axis and another out-of-plane direction. Though these two band structures are very similar over most of the Brillouin zone, the slope of the flat band crossing the Fermi energy along $\Gamma - K$ and $\Gamma - M$ are found to be opposite. The calculated band structure with magnetization along the in-plane direction is in agreement with Fig. 3(a) of the main text, which is obtained from the VASP calculations.

### E. Calculation of magnetic interaction parameters

The hopping integrals for the non-spin-polarized $\alpha$-RuCl$_3$/gr heterostructure shown in Table III, were obtained using the Wannier function projection formalism proposed in Refs. 7–9. The *ab initio* density functional theory (DFT) calculations are performed with the full-potential-linearized-augmented-plane-wave basis (LAPW) method as implemented in WIEN2k [10]. The Perdew-Burke-Ernzerhof generalized gradient approximation [11] was used, with a mesh of 1000 **k** points in the first Brillouin zone and RK$_{\text{max}}$ was set to 8. In Table III, the values of the hopping parameters for $\alpha$-RuCl$_3$/gr are compared to those of bulk $\alpha$-RuCl$_3$ from Ref. [12].

We note that the $t - J$ model described in Eq. 1 of the main text, is expressed in terms of relativistic $j_{\text{eff}}$ basis. However, in table III, the hopping matrices are defined in terms of the non-relativistic $t_{2g}$ basis. The hopping parameters for the Z-bond (nearest neighbor) in the strained $\alpha$-RuCl$_3$ is given by [12]:

$$T^Z = \begin{pmatrix} t_1 & t_2 & t_4 \\ t_2 & t_1 & t_4 \\ t_4 & t_4 & t_3 \end{pmatrix} \quad (1)$$

The $T^Z$ matrix above is defined in the $C_{t_{2g}} = \{c_{yz,\uparrow}, c_{yz,\downarrow}, c_{xz,\uparrow}, c_{xz,\downarrow}, c_{xy,\uparrow}, c_{xy,\downarrow}\}$ basis and the hopping parameters $t_i$ for i=1-4 are defined in the same way as in Ref. [12]. In order to transform the $C_{t_{2g}}$ basis to the relativistic basis of $C_{j_{\text{eff}}}$, we can represent the $T^Z$ matrix in the $j_{\text{eff}} = \{\{j, m_j\}\} =$

Table III: Parameters for crystal field splitting and nearest neighbor hopping (meV) for monolayer and experimental C2/m structures of RuCl$_3$

| Term | $\alpha$-RuCl$_3$/gr | $C/2m$ Ref. [12] |
|---|---|---|
| $\Delta_1$ | -8.4 | -19.8 |
| $\Delta_2$ | -8.7 | -17.5 |
| $\Delta_3$ | -0.1 | -12.5 |
| $t_1$ ($t_{\bar{1}\|}$) | +22.7 | +50.9 |
| $t'_{1a}$ ($t_{1\|}$) | +17.3 | +44.9 |
| $t'_{1b}$ ($t_{1\|}$) | +26.5 | +45.8 |
| $t_2$ ($t_{\bar{1}O}$) | +176.1 | +158.2 |
| $t'_2$ ($t_{1O}$) | +175.1 | +162.2 |
| $t_3$ ($t_{\bar{1}\sigma}$) | -24.1 | -154.0 |
| $t'_3$ ($t_{1\sigma}$) | -22.3 | -103.1 |
| $t_4$ ($t_{\bar{1}\perp}$) | -16.6 | -20.2 |
| $t'_{4a}$ ($t_{1\perp}$) | -21.9 | -15.1 |
| $t'_{4b}$ ($t_{1\perp}$) | -20.8 | -10.9 |

$\{\{\frac{1}{2}, \frac{1}{2}\}, \{\frac{1}{2}, -\frac{1}{2}\}, \{\frac{3}{2}, \frac{3}{2}\}, \{\frac{3}{2}, \frac{1}{2}\}, \{\frac{3}{2}, -\frac{1}{2}\}, \{\frac{3}{2}, -\frac{3}{2}\}\}$ basis. For this, we have used the following transformation matrix:

$$\begin{pmatrix} \frac{1}{\sqrt{3}} & 0 & \frac{-i}{\sqrt{3}} & 0 & 0 & \frac{-1}{\sqrt{3}} \\ 0 & \frac{1}{\sqrt{3}} & 0 & \frac{i}{\sqrt{3}} & \frac{1}{\sqrt{3}} & 0 \\ 0 & \frac{1}{\sqrt{2}} & 0 & \frac{-i}{\sqrt{2}} & 0 & 0 \\ 0 & \frac{-1}{\sqrt{6}} & 0 & \frac{-i}{\sqrt{6}} & \frac{\sqrt{2}}{\sqrt{3}} & 0 \\ \frac{1}{\sqrt{6}} & 0 & \frac{-i}{\sqrt{6}} & 0 & 0 & \frac{\sqrt{2}}{\sqrt{3}} \\ \frac{-1}{\sqrt{2}} & 0 & \frac{-i}{\sqrt{2}} & 0 & 0 & 0 \end{pmatrix} \quad (2)$$

Then the transformed matrix $T^Z_{j_{\text{eff}}}$, is given by the following equation:

$$\begin{pmatrix} \frac{2t_1+t_3}{3} & 0 & -\frac{(1+i)t_4}{\sqrt{6}} & \frac{(1-i)t_4}{\sqrt{2}} & \frac{\sqrt{2}(t_1-t_3)}{3} & i\sqrt{\frac{2}{3}}t_2 \\ 0 & \frac{2t_1+t_3}{3} & i\sqrt{\frac{2}{3}}t_2 & \frac{\sqrt{2}(-t_1+t_3)}{3} & \frac{(1+i)t_4}{\sqrt{2}} & -\frac{(1-i)t_4}{\sqrt{6}} \\ -\frac{(1-i)t_4}{\sqrt{6}} & -i\sqrt{\frac{2}{3}}t_2 & t_1 & \frac{it_2}{\sqrt{3}} & \frac{(1-i)t_4}{\sqrt{3}} & 0 \\ \frac{(1+i)t_4}{\sqrt{2}} & \frac{\sqrt{2}(-t_1+t_3)}{3} & \frac{-it_2}{\sqrt{3}} & \frac{t_1+2t_3}{3} & 0 & -\frac{(1-i)t_4}{\sqrt{3}} \\ \frac{\sqrt{2}(t_1-t_3)}{3} & \frac{(1-i)t_4}{\sqrt{2}} & \frac{(1+i)t_4}{\sqrt{3}} & 0 & \frac{t_1+2t_3}{3} & \frac{it_2}{\sqrt{3}} \\ -i\sqrt{\frac{2}{3}}t_2 & -\frac{(1+i)t_4}{\sqrt{6}} & 0 & -\frac{(1+i)t_4}{\sqrt{3}} & -i\frac{t_2}{\sqrt{3}} & t_1 \end{pmatrix} \quad (3)$$

Similarly, the hopping matrices for the X and Y bonds (nearest neighbor), which are given by the following equation in the $t_{2g}$ basis,

$$T^{\text{X}} = \begin{pmatrix} t'_3 & t'_{4a} & t'_{4b} \\ t'_{4a} & t'_{1a} & t'_2 \\ t'_{4b} & t'_2 & t'_{4b} \end{pmatrix}, T^{\text{Y}} = \begin{pmatrix} t'_{1a} & t'_{4a} & t'_2 \\ t'_{4a} & t'_3 & t'_{4b} \\ t'_2 & t'_{4b} & t'_{1b} \end{pmatrix} \quad (4)$$

transform in the following way:

$$\begin{pmatrix} \frac{t'_{1a}+t'_{1b}+t'_3}{3} & 0 & -i\frac{(t'_2-it'_{4b})}{\sqrt{6}} & \frac{-it'_2+t'_{4b}}{\sqrt{2}} & \frac{t'_{1a}-2t'_{1b}+t'_3}{3\sqrt{2}} & \frac{t'_{1a}-t'_3+2it'_{4a}}{\sqrt{6}} \\ 0 & \frac{t'_{1a}+t'_{1b}+t'_3}{3} & \frac{-t'_{1a}+t'_3+2it'_{4a}}{\sqrt{6}} & -\frac{(t'_{1a}-2t'_{1b}+t'_3)}{3\sqrt{2}} & \frac{it'_2+t'_{4b}}{\sqrt{2}} & i\frac{(t'_2+it'_{4b})}{\sqrt{6}} \\ i\frac{t'_2+it'_{4b}}{\sqrt{6}} & \frac{-t'_{1a}+t'_3-2it'_{4a}}{\sqrt{6}} & \frac{t'_{1a}+t'_3}{2} & \frac{t'_{1a}-t'_3+2it'_{4a}}{2\sqrt{3}} & \frac{-it'_2+t'_{4b}}{\sqrt{2}} & 0 \\ \frac{it'_2+t'_{4b}}{\sqrt{2}} & -\frac{(t'_{1a}-2t'_{1b}+t'_3)}{3\sqrt{2}} & \frac{t'_{1a}-t'_3-2it'_{4a}}{2\sqrt{3}} & \frac{t'_{1a}+4t'_{1b}+t'_3}{6} & 0 & i\frac{(t'_2+it'_{4b})}{\sqrt{3}} \\ \frac{t'_{1a}-2t'_{1b}+t'_3}{3\sqrt{2}} & \frac{-it'_2+t'_{4b}}{\sqrt{2}} & \frac{it'_2+t'_{4b}}{\sqrt{3}} & 0 & \frac{t'_{1a}+4t'_{1b}+t'_3}{6} & \frac{t'_{1a}-t'_3+2it'_{4a}}{2\sqrt{3}} \\ \frac{t'_{1a}-t'_3-2it'_{4a}}{\sqrt{6}} & \frac{-it'_2-t'_{4b}}{\sqrt{6}} & 0 & \frac{-it'_2-t'_{4b}}{\sqrt{3}} & \frac{t'_{1a}-t'_3-2it'_{4a}}{2\sqrt{3}} & \frac{t'_{1a}+t'_3}{2} \end{pmatrix}$$

and

$$\begin{pmatrix} \frac{t'_{1a}+t'_{1b}+t'_3}{3} & 0 & -\frac{(t'_2+it'_{4b})}{\sqrt{6}} & \frac{t'_2-it'_{4b}}{\sqrt{2}} & \frac{t'_{1a}-2t'_{1b}+t'_3}{3\sqrt{2}} & \frac{-t'_{1a}+t'_3+2it'_{4a}}{\sqrt{6}} \\ 0 & \frac{t'_{1a}+t'_{1b}+t'_3}{3} & \frac{t'_{1a}-t'_3+2it'_{4a}}{\sqrt{6}} & -\frac{(t'_{1a}-2t'_{1b}+t'_3)}{3\sqrt{2}} & \frac{t'_2+it'_{4b}}{\sqrt{2}} & -\frac{(t'_2-it'_{4b})}{\sqrt{6}} \\ -\frac{(t'_2-it'_{4b})}{\sqrt{6}} & \frac{t'_{1a}-t'_3-2it'_{4a}}{\sqrt{6}} & \frac{t'_{1a}+t'_3}{2} & \frac{-t'_{1a}+t'_3+2it'_{4a}}{2\sqrt{3}} & \frac{t'_2-it'_{4b}}{\sqrt{3}} & 0 \\ \frac{t'_2+it'_{4b}}{\sqrt{2}} & -\frac{(t'_{1a}-2t'_{1b}+t'_3)}{3\sqrt{2}} & -\frac{(t'_{1a}-t'_3-2it'_{4a})}{2\sqrt{3}} & \frac{t'_{1a}+4t'_{1b}+t'_3}{6} & 0 & -\frac{(t'_2-it'_{4b})}{\sqrt{3}} \\ \frac{t'_{1a}-2t'_{1b}+t'_3}{3\sqrt{2}} & \frac{t'_2-it'_{4b}}{\sqrt{2}} & \frac{t'_2-it'_{4b}}{\sqrt{3}} & 0 & \frac{t'_{1a}+4t'_{1b}+t'_3}{6} & \frac{-t'_{1a}+t'_3+2it'_{4a}}{2\sqrt{3}} \\ \frac{-t'_{1a}+t'_3-2it'_{4a}}{\sqrt{6}} & -\frac{(t'_2+it'_{4b})}{\sqrt{6}} & 0 & -\frac{(t'_2+it'_{4b})}{\sqrt{3}} & -\frac{(t'_{1a}-t'_3+2it'_{4a})}{2\sqrt{3}} & \frac{t'_{1a}+t'_3}{2} \end{pmatrix}$$

(5)


\clearpage

\end{document}